\documentclass{article}
\usepackage{booktabs}
\usepackage{CJK}
\usepackage{comment}
\usepackage{amssymb} % 导入 amssymb 宏包

\usepackage{subcaption}
\usepackage[english]{babel}

\usepackage[letterpaper,top=2cm,bottom=2cm,left=3cm,right=3cm,marginparwidth=1.75cm]{geometry}

\usepackage{amsmath}
\usepackage{graphicx}
\usepackage{authblk}
\usepackage[colorlinks=true, allcolors=blue]{hyperref}

\title{Introducing voice timbre attribute detection}

\author{
  \begin{center}
    Jinghao He$^{1}$, Zhengyan Sheng$^{1}$, Liping Chen$^{1}$, Kong Aik Lee$^{2}$, Zhen-Hua Ling$^{1}$ \\
    $^{1}$NERC-SLIP, University of Science and Technology of China, China \\
    $^{2}$Department of Electrical and Electronic Engineering, The Hong Kong Polytechnic University, Hong Kong \\
    \texttt{\{jhhe, zysheng\}@mail.ustc.edu.cn}\\
    \texttt{\{lipchen, zhling\}@ustc.edu.cn} \\
    \texttt{kong-aik.lee@polyu.edu.hk}
  \end{center}
}

\begin{document}
\maketitle

%%%%  摘要   %%%%
\begin{abstract}
This paper focuses on explaining the timbre conveyed by speech signals and introduces a task termed voice timbre attribute detection (vTAD). In this task, voice timbre is explained with a set of sensory attributes describing its human perception. A pair of speech utterances is processed, and their intensity is compared in a designated timbre descriptor. Moreover, a framework is proposed, which is built upon the speaker embeddings extracted from the speech utterances. The investigation is conducted on the VCTK-RVA dataset. Experimental examinations on the ECAPA-TDNN and FACodec speaker encoders demonstrated that: 1) the ECAPA-TDNN speaker encoder was more capable in the seen scenario, where the testing speakers were included in the training set; 2) the FACodec speaker encoder was superior in the unseen scenario, where the testing speakers were not part of the training, indicating enhanced generalization capability. The VCTK-RVA dataset and open-source code are available on the website \url{https://github.com/vTAD2025-Challenge/vTAD}.  

\end{abstract}
%
%

% 第一部分 介绍
\section{Introduction}
%Timbre, as a fundamental aspect of hearing, plays an important role in speech perception. 

As a crucial component of the information conveyed by the speech signal, voice attributes can be perceived by both machine algorithms and human hearing. In the past years, algorithms for voice attributes modeling have been widely investigated, achieving remarkable advancements, and significantly promoting speech technologies, including speaker recognition \cite{x-vector,desplanques2020ecapa,ExplainableASV}, speech recognition \cite{li2022recent}, speech generation \cite{yourtts,voice_conversion_overview}, etc. Recently, human perception of voice attributes, known as timbre, has attracted the attention of the research community. To name a few, studies of the asynchronous voice anonymization techniques \cite{wang2024async} have revealed the inconsistency between human and machine perceptions of voice attributes in speech signals. The voice editing method introduced in \cite{DBLP:journals/corr/abs-2404-08857} facilitated modifying the timbre of the voice in speech utterances.

%Voice characteristics, which express the speaker's identity, are a crucial component of speech. Voice characteristics exists at the confluence of the physical and the perceptual \cite{wallmark2018describing}. We often encroach upon the domains of sight, feeling, and even taste \cite{wallmark2019corpus}, using terms like bright, sharp, and dry to express our perception of voice characteristics.

To achieve a further understanding of timbre, this paper focuses on its explainability and aims to uncover the relationship between speech acoustics and timbre. Specifically, the human impressions of timbre are verbalized, based on which the timbre attributes are explained. To this end, a set of timbre descriptors derived from sensory attributes across various modalities is built, including sound (hoarse, rich), vision (bright, dark), texture (soft, hard), physical attribute (magnetic, transparent), and so on. Based on that, a novel task is defined, namely voice timbre attribute detection (vTAD). In this task, given a pair of speech utterances, the intensity difference between them in a specific descriptor dimension is detected, and the comparison outcome is obtained.

In this paper, a framework is proposed for vTAD, built upon the speaker embedding vector. Given the speaker embedding vectors derived from the utterance pair, a neural network is employed to compare the timbre attributes and predict the intensity difference across the descriptor dimensions. In our study, two speaker embedding vector extractors are examined, including ECAPA-TDNN \cite{desplanques2020ecapa}and FACodec\cite{facodec}. Experiments conducted on the VCTK-RVA dataset \cite{vctk-rva} demonstrate that: 1) in the unseen scenario, where the evaluated speakers were not included in the training set, the FACodec speaker embedding extractor exhibited better generalization capabilities, and 2) in the seen scenario, where the evaluated speakers were included in the training set, the ECAPA-TDNN speaker extractor achieved superior performances.

\section{Task Definition}

\begin{figure}[htbp]
\centering
\includegraphics[scale=1.0]{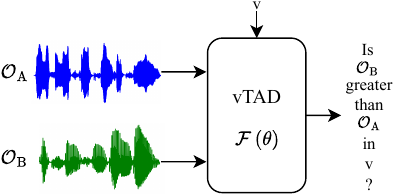}
\caption{Task definition of vTAD.}
\label{fig: task definition}
\end{figure}

In this task, a timbre attribute descriptor set is defined as ${\mathcal V}$. As shown in Fig. \ref{fig: task definition}, given a pair of utterances ${\mathcal O}_{\rm A}$ and ${\mathcal O}_{\rm B}$ from speakers A and B, respectively, and a designated timbre descriptor v$\in\mathcal{V}$, the vTAD evaluates whether the intensity of v in ${\mathcal O}_{\rm B}$ is stronger than that in ${\mathcal O}_{\rm A}$. Mathematically, the hypothesis about the intensity difference is defined as ${\mathcal H}\left(\langle{\mathcal O}_{\rm A}, {\mathcal O}_{\rm B}\rangle, {\rm v}\right)$, meaning that ${\mathcal O}_{\rm B}$ is stronger than ${\mathcal O}_{\rm A}$ in the descriptor dimension v. Specifically, ${\mathcal H} \in \{0, 1\}$, where ${\mathcal H} = 1$ indicates that the hypothesis ${\mathcal H}$ is correct, and ${\mathcal H} = 0$ indicates that the hypothesis is incorrect. The hypothesis is determined by the vTAD algorithm function ${\mathcal F}\left(\langle{\mathcal O}_{\rm A}, {\mathcal O}_{\rm B}\rangle | {\rm v}; \theta\right)$, with $\theta$ representing the set of algorithm parameters.

\section{Dataset Description}

The VCTK-RVA dataset \cite{vctk-rva} is employed in our work, wherein the publicly available VCTK database was annotated for timbre intensity. In the dataset, 18 timbre descriptors are defined in $\mathcal V$, as listed in Table \ref{table1}. In total, 101 speakers are involved, forming 6,038 annotated ordered speaker pairs \{Speaker A, Speaker B, voice attribute v\}, indicating that Speaker B is stronger than Speaker A in the specific descriptor v. The number of descriptor dimensions annotated for each ordered speaker pair ranges from 1 to 3.
%Only the most salient voice attribute are annotated; the absence of annotation for certain attributes does not imply the absence of differences. 

% In this paper, the VCTK-RVA dataset is partitioned into one training set, a seen speaker test set, an unseen speaker test set and a seen annotation test set. Specifically, the training set contains 78 speakers and 18 voice attributes. All speakers in the seen speaker test set also appear in the training set, but the annotated groups do not overlap with those in the training set. In contrast, the speakers in the unseen speaker test set do not appear in the training set at all.The annotation groups of the seen annotation test set completely overlap with those of the training set, while the selected utterances do not overlap with those in the training set at all. Data statistics for the training set, seen speaker test set, and unseen speaker test set are presented in Table \ref{table2}, Table \ref{table3}, and Table \ref{table4}, respectively.

% table1 统计attribute频率
\begin{table}[h]
	\centering
		\caption{The descriptor set used for describing the timbre. The \emph{Trans.} column gives the corresponding Chinese word. The \emph{Perc.} column presents the percentage (\%) of each descriptor in the \emph{VCTK-RVA} dataset. The descriptors shrill and husky are exclusively annotated for female and male, respectively.}
	\begin{tabular}{lcc lcc}
		\toprule[1pt]
		
		\textbf{Descriptor} & \textbf{Trans.} & \textbf{Perc.} & \textbf{Descriptor} & \textbf{Trans.} & \textbf{Perc.} \\ 
		\cmidrule(lr){1-3} \cmidrule(lr){4-6}
		Bright      & \begin{CJK*}{UTF8}{gbsn}明亮\end{CJK*}      & 17.10      & Thin  &\begin{CJK*}{UTF8}{gbsn}单薄\end{CJK*}     & 13.03      \\
		Coarse &\begin{CJK*}{UTF8}{gbsn}粗\end{CJK*}      & 11.62      & Slim  &\begin{CJK*}{UTF8}{gbsn}细\end{CJK*}      & 11.31      \\
		Low &\begin{CJK*}{UTF8}{gbsn}低沉\end{CJK*}        & 7.43       & Pure &\begin{CJK*}{UTF8}{gbsn}干净\end{CJK*}       & 5.48       \\
		Rich &\begin{CJK*}{UTF8}{gbsn}厚实\end{CJK*}       & 4.71       & Magnetic  &\begin{CJK*}{UTF8}{gbsn}磁性\end{CJK*}  & 3.64       \\
		Muddy  &\begin{CJK*}{UTF8}{gbsn}浑浊\end{CJK*}     & 3.59       & Hoarse &\begin{CJK*}{UTF8}{gbsn}沙哑\end{CJK*}     & 3.32       \\
		Round  &\begin{CJK*}{UTF8}{gbsn}圆润\end{CJK*}     & 2.48       & Flat  &\begin{CJK*}{UTF8}{gbsn}平淡\end{CJK*}      & 2.15       \\
		Shrill  (female only) &\begin{CJK*}{UTF8}{gbsn}尖锐\end{CJK*}     & 2.08       & Shriveled &\begin{CJK*}{UTF8}{gbsn}干瘪\end{CJK*}  & 1.74       \\
		Muffled &\begin{CJK*}{UTF8}{gbsn}沉闷\end{CJK*}    & 1.44       & Soft    &\begin{CJK*}{UTF8}{gbsn}柔和\end{CJK*}    & 0.82       \\
		Transparent &\begin{CJK*}{UTF8}{gbsn}通透\end{CJK*} & 0.66       & Husky (male only)&\begin{CJK*}{UTF8}{gbsn}干哑\end{CJK*}      & 0.59       \\ 
		\bottomrule[1pt]
	\end{tabular}

	\label{table1}
\end{table}

\section{Method}
% order samples

Given an annotated utterance pair \(\left\{\langle{\mathcal O}_{\rm A},{\mathcal O}_{\rm B}\rangle, {\rm v}\right\}\) indicating that $\mathcal{O}_{\rm B}$ is stronger than $\mathcal{O}_{\rm A}$ in the descriptor dimension v, a training sample is obtained as \(\left\{\langle{\mathcal O}_{\rm A},{\mathcal O}_{\rm B}\rangle, {\boldsymbol{l}}\right\}\) where ${\boldsymbol{l}}$ is the ground-truth label vector. Assuming that the timbre attribute descriptor set ${\mathcal V}$ is composed of \( N \) descriptors, \({\boldsymbol{l}}\) is an \( N \)-dimensional vector, where the \( n \)-th dimension $l_{n}$ represents the comparative intensity of \({\mathcal O}_{\rm A}\) and \({\mathcal O}_{\rm B}\) for the \( n \)-th descriptor with $l_{n}\in\left\{0,1,-1\right\}$. Here, \( l_n = 1 \) indicates that in the $n$-th descriptor dimension, ${\mathcal O}_{\rm B}$ is stronger than ${\mathcal O}_{\rm A}$; \( l_n = 0 \) indicates that ${\mathcal O}_{\rm B}$ is not stronger than ${\mathcal O}_{\rm A}$; and \( l_n = -1 \) indicates that no intensity comparison is applied between ${\mathcal O}_{\rm B}$ and ${\mathcal O}_{\rm A}$ in the $n$-th descriptor. The proposed model is illustrated in Fig. \ref{fig: model architecture}.
% y
%By combining the labels \( \mathit{l} \) from the $N$ descriptor dimensions v,we obtain a $N$-dimensional true label vector $y$,which encapsulates the comparison between ${\mathcal O}_{\rm A}$ and ${\mathcal O}_{\rm B}$ across $N$ descriptor dimensions v.

%it presents a detailed structure of vTAD models constructed using different pre-trained speaker encoders,such as ECAPA-TDNN\cite{DBLP:conf/interspeech/DesplanquesTD20}, FACodec\cite{DBLP:conf/icml/JuWS0XYLLST000024}, Xi-Vector, ResNet, and so on.

% e^A/e^B
\subsection{Overall Architecture}
\begin{figure}[t]
\centering
\begin{subfigure}[t]{1\textwidth}
        \centering
        \includegraphics[scale=1.0]{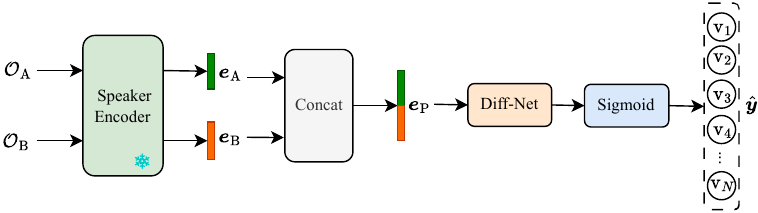}
        \caption{Overall architecture}
        \label{fig: overall architecture}
    \end{subfigure}
    \hfill
    \vspace{0.4cm}
    \begin{subfigure}[t]{1\textwidth}
        \centering
        \includegraphics[scale=1.0]{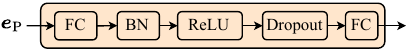}
        \caption{Diff-Net}
        \label{fig: diff-net}
    \end{subfigure}
    \caption{Proposed framework for vTAD. The speaker encoder is pre-trained and frozen. The notation Conca is short for concatenation, FC is short for fully-connected layer, BN is short for batch normalization.}
\label{fig: model architecture}
\end{figure}

As depicted in Fig.\ref{fig: model architecture}(\subref{fig: overall architecture}), in the proposed framework, speaker embedding vectors are extracted from the utterance pair ${\mathcal O}_{\rm A}$ and ${\mathcal O}_{\rm B}$ with a speaker encoder, and represented with ${\boldsymbol{e}}_{\rm A}$ and ${\boldsymbol{e}}_{\rm B}$, respectively. The concatenation of both is obtained as $e_{\rm P}$, which is then input to the Diff-Net module. The output dimension of the Diff-Net is $N$. Thereafter, the sigmoid function is applied to each node, producing the vector \({\hat{\boldsymbol{y}}}\), where the \(n\)-th $\left(n=1,2,...,N\right)$ dimension is the prediction of the intensity comparison for the corresponding descriptor. In this framework, the speaker encoder is pre-trained and frozen, which can be derived from speaker models such as x-vector \cite{x-vector}, ECAPA-TDNN \cite{desplanques2020ecapa}, ResNet \cite{resnet}, FACodec \cite{facodec}, etc.

% \paragraph{ECAPA-TDNN:}Pre-trained on the VoxCeleb1 and VoxCeleb2 datasets. The core idea of ECAPA-TDNN is to enhance the representational capacity of speaker features by optimizing the model architecture, with a particular focus on improving performance in extracting vocal characteristics from speech.

% \paragraph{FACodec:}FACodec adopts a hierarchical VQ-VAE architecture, combined with multi-scale spectral adversarial training and a dynamic loss balancing mechanism, enabling stable end-to-end optimization on a large-scale dataset of 50k hours.

% \paragraph{PairEmbedding:}In Figure \ref{figure2},this module generates a 34-dimensional attribute ground truth label vector y according to different concatenation patterns.For the positions corresponding to the annotated attributes, they are assigned values of either 0 or 1. If a specific voice attribute is not annotated, the corresponding position in the vector is assigned a value of -1 and is excluded from the subsequent loss computation.

% \paragraph{Discriminator:}In Figure \ref{figure2}(c), the module classifies the concatenated embedding representation \( S \) using multiple fully connected layers. After passing through a Sigmoid activation function, it generates a 34-dimensional predicted label \(\hat{y}\). Each dimension of \(\hat{y}\) corresponds to the classification result of a specific attribute. Since there are 17 attributes for each gender, only the 17 attribute classifications corresponding to the speaker's gender are considered.

\subsection{Diff-Net}
As shown in Fig. \ref{fig: model architecture}(\subref{fig: diff-net}), the Diff-Net takes the concatenated embedding ${\boldsymbol{e}}_{\rm P}$ as input. A fully-connected layer is applied with a batch normalization layer. ReLU is adopted as the activation function, and dropout is applied. Then, another fully-connected layer is applied with the output dimension of $N$.

\subsection{Loss Function} During model training, only the labeled descriptor, $l_n\in\left\{0,1\right\}$, is accounted for in loss computation. With this, given an utterance pair $\langle{\mathcal O}_{\rm A},{\mathcal O}_{\rm B} \rangle$, only the labeled descriptor dimension is used in model training. The loss function is defined as:

\begin{equation}
    {\mathcal L} =  \mathbb{I}[l_{n}\in\left\{0,1\right\}]\cdot BCE\left(l_{n}, \hat{y}_{n}\right), 
\end{equation}
where $BCE\left(\bullet\right)$ is the binary cross-entropy function computed as follows:
\begin{equation}
    BCE(l_{n}, \hat{y}_{n}) = - l_{n}log(\hat{y}_{n}) - (1-l_{n})log(1-\hat{y}_{n}).
\end{equation}
The loss function minimizes the cross-entropy between the prediction and the ground-truth, ensuring that the model can accurately predict the intensity difference across the descriptor dimensions of each utterance pair.

\subsection{Inference}
During inference, given an ordered utterance pair \(\langle{\mathcal O}_{\rm A},{\mathcal O}_{\rm B}\rangle\) and a specific descriptor \(v\), speaker embedding vectors are extracted from both utterances. After being processed by the Diff-Net, the output vector of dimension $N$ is generated and then processed by the sigmoid function to obtain the vector ${\hat {\boldsymbol{y}}}$. Each dimension in ${\hat {\boldsymbol{y}}}$ corresponds to a descriptor in $\mathcal V$. Particularly, the output value of the node corresponding to the designated descriptor v is obtained and utilized as the confidence score quantifying the likelihood that \({\mathcal O}_{\rm B}\) is stronger than \({\mathcal O}_{\rm A}\) in the descriptor dimension v.

\begin{figure}[t]
\centering
\includegraphics[scale=0.6]{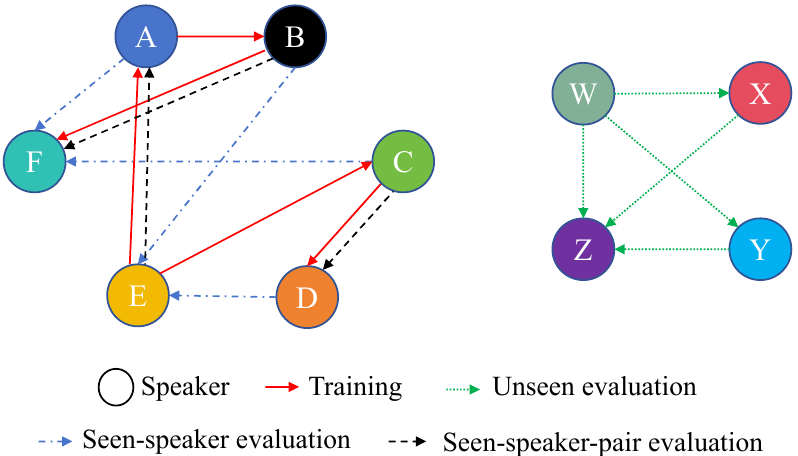}
\caption{Ordered speaker pair construction in training, unseen, seen-speaker, and seen-speaker-pair evaluations, respectively. The direction of the arrow is from the weaker speaker to the stronger speaker in a specific descriptor in the training annotation and evaluation hypothesis.}
\label{fig: data_pairs}
\end{figure}

\section{Evaluation}
\subsection{Metrics}
Evaluations are conducted on speech utterances ${\mathcal O}_{\rm A}$ and $\mathcal{O}_{\rm B}$, originating from a pair of speakers A and B, respectively. The performance is evaluated in two tasks: verification and recognition. The hypothesis \({\mathcal{H}\left(\langle{\mathcal O}_{\rm A},{\mathcal O}_{\rm B}\rangle, {\rm v}\right)}=1\), where ${\rm v}\in {\mathcal V}$, is defined, assuming that ${\mathcal O}_{\rm B}$ is stronger than $\mathcal{O}_{\rm A}$ in the descriptor dimension v. The system provides the confidence score of \({\mathcal H}\) in the verification evaluation and determines whether \({\mathcal H}\) is correct in the recognition evaluation. The verification results are measured with equal error rate (EER), and the recognition results are measured with accuracy. The lower EERs and higher accuracies indicate better performance.

\begin{itemize}
\item \emph{EER}: In the verification evaluation, the target and nontarget trials are composed regarding whether the hypothesis \({\mathcal{H}\left(\langle{\mathcal O}_{\rm A},{\mathcal O}_{\rm B}\rangle, {\rm v}\right)}=1\) is true or not. Specifically, the target evaluation samples consist of instances where \({\mathcal{H}\left(\langle{\mathcal O}_{\rm A},{\mathcal O}_{\rm B}\rangle, {\rm v}\right)}=1\), while the nontarget samples comprise instances where \({\mathcal{H}\left(\langle{\mathcal O}_{\rm A},{\mathcal O}_{\rm B}\rangle, {\rm v}\right)}=0\). Given an evaluation sample $\left\{\langle{\mathcal O}_{\rm A},{\mathcal O}_{\rm B}\rangle, {\rm v}\right\}$, denote the confidence score obtained by the algorithm as $s_{\langle{\rm A},{\rm B}\rangle}^{\rm v}$. Higher $s_{\langle{\rm A},{\rm B}\rangle}^{\rm v}$ value indicates that ${\mathcal O}_{\rm B}$ is more likely to be stronger than ${\mathcal O}_{\rm A}$ in the descriptor dimension v. Finally, the EER value is computed on the confidence scores given the ground-truth target and nontarget labels of the evaluations samples.

\item \emph{Accuracy (ACC)}:
In the recognition evaluation, given the evaluation sample \(\left\{\langle{\mathcal O}_{\rm A},{\mathcal O}_{\rm B}\rangle,{\rm v}\right\}\) and the ground-truth label \(t \in \{0, 1\}\), a label of 0 indicates that the hypothesis \({\mathcal{H}\left(\langle{\mathcal O}_{\rm A},{\mathcal O}_{\rm B}\rangle, {\rm v}\right)}=1\) is false, while a label of 1 indicates that the hypothesis is true. The algorithm predicts whether the hypothesis ${\mathcal H}$ is true or not. Thereby, the accuracy is computed between the prediction and the ground truth $t$ as follows:

\begin{equation}
\text{ACC} = \frac{\text{TP} + \text{TN}}{\text{TP} + \text{TN} + \text{FP} + \text{FN}}.
\label{eq: ACC}
\end{equation}
In (\ref{eq: ACC}), TP is short for true positives, representing the number of true evaluation samples that are correctly predicted. TN is short for true negatives, representing the number of false evaluation samples that are correctly predicted. FP is short for false positives, denoting the number of false evaluation samples that are incorrectly predicted to be true. FN is short for false negative, denoting the number of true evaluation samples that are incorrectly predicted to be false.

\end{itemize}

\noindent For both EER and ACC, the results obtained by averaging across all evaluated descriptors are used as the metrics of system performance.

\subsection{Scenarios}
\label{sec:eval scenarios}
As depicted in Fig. \ref{fig: data_pairs}, regarding the speakers applied in the training and evaluation, three evaluation scenarios are defined, including unseen, seen-speaker, and seen-speaker-pair. The detailed descriptions are as follows.

\begin{itemize}
    \item \emph{Unseen}: In the unseen scenario, the speakers used in the evaluation phase are not present in the training phase.
    \item \emph{Seen-speaker}: In this scenario, the speakers employed in the evaluation phase are applied in the training phase, while distinct utterances are utilized for training and evaluation, respectively. Moreover, given a specific speaker, the ordered pairs composed with different speakers are exclusively used in training and evaluation.
    \item \emph{Seen-speaker-pair}: In this scenario, the ordered speaker pairs utilized in the evaluation phase are also included in the training phase; however, different utterances are employed for each speaker in the training and evaluation phases, respectively.
\end{itemize}

\subsection{Data Splits}
In this challenge, the VCTK-RVA dataset is partitioned for training and evaluation, respectively. For each gender, the training set contains speaker pairs annotated on all 17 descriptors. In total, 29 male and 49 female speakers are included in the training phase. In the unseen and seen-speaker evaluations, five descriptors are selected for each gender. The seen-speaker-pair evaluations were conducted on all 17 descriptors for each gender. Data statistics for the training set, unseen, and seen-speaker test sets are presented in Table \ref{tab: training spkr stats}, \ref{tab: unseen-spkr eval stats}, \ref{tab: seen-spkr eval stats}, respectively. The seen-speaker-pair evaluation was built upon all the speaker pairs in the training set. The evaluation trials are configured as follows.

\begin{itemize}
    \item \emph{Unseen}: In each ordered speaker pair, 20 speech utterances are randomly selected for each speaker, resulting in 400 utterance pairs per ordered speaker pair. This leads to a total of 91,600 utterance pairs in the unseen evaluation set.
    \item \emph{Seen-speaker}: In each ordered speaker pair, 20 speech utterances are randomly selected for each speaker, resulting in 400 utterance pairs per ordered speaker pair. This leads to a total of 94,000 utterance pairs in the unseen evaluation set.
    % \item \emph{Seen-speaker}: Similar to the number of utterances selected in the unseen scenario, each ordered speaker pair generates 400 utterance pairs, resulting in a total of 94,000 utterance pairs for the seen speaker set.
     \item \emph{Seen-speaker-pair}: In each ordered speaker pair, 10 speech utterances are randomly selected for each speaker, resulting in 100 utterance pairs per ordered speaker pair. This leads to a total of 340,800 utterance pairs in the unseen evaluation set.
    % \item \emph{Seen-speaker-pair}: In the seen-speaker-pair scenario, for each speaker in every ordered speaker pair, 10 speech samples are selected, resulting in 100 utterance pairs per ordered speaker pair. This leads to a total of 340,800 utterance pairs in the seen-speaker-pair set.
\end{itemize}

\begin{comment}

\subsection{Evaluation Metrics}
When evaluating the performance of a classification model, common metrics include accuracy (ACC) and equal error rate (EER). This paper selects these two objective metrics as the evaluation criteria for the vTAD model. Accuracy is defined as the ratio of correctly classified samples to the total number of samples, as shown in the following formula:

\begin{equation}
\text{ACC} = \frac{\text{TP} + \text{TN}}{\text{TP} + \text{TN} + \text{FP} + \text{FN}}
\end{equation}

where TP denotes true positives, TN denotes true negatives, FP denotes false positives, and FN denotes false negatives. Accuracy reflects the model's classification capability on the entire dataset.

However, when the dataset is imbalanced, relying solely on accuracy can be misleading. In this case, the equal error rate (EER) is a commonly used evaluation metric, particularly in tasks such as authentication. The EER is the error rate when the false positive rate (FPR) and false negative rate (FNR) are equal at different decision thresholds. The formulas for calculating the false positive rate and false negative rate are as follows:

\begin{equation}
\text{FPR} = \frac{\text{FP}}{\text{FP} + \text{TN}}
\end{equation}

\begin{equation}
\text{FNR} = \frac{\text{FN}}{\text{FN} + \text{TP}}
\end{equation}

The EER point is the threshold where the FPR and FNR are equal. At the EER, the model achieves a balance between precision and recall, thus providing a more comprehensive reflection of the model's actual performance. 

\end{comment}

\subsection{Configurations}

As the VCTK-RVA dataset specifically annotates significant differences in voice attributes among speakers of the same gender, the descriptor dimensions v require gender-specific processing during training and testing. As shown in Table \ref{table1}, each of the male and female categories is annotated with 17 voice attributes, resulting in a total of 34 descriptor dimensions. Our proposed model was trained in a gender-dependent manner, with separate training targets for females and males, resulting in \( N \) being set to 34. During the training, all attributes listed in Table \ref{table1} were used. Additionally, all speech samples were downsampled to 16 kHz. For each ordered speaker pair, 20 utterances were randomly selected from each speaker to create the training samples.

Two pre-trained speaker encoders were experimented with for speaker embedding extraction: ECAPA-TDNN and FACodec. In the FACodec model, the timbre extractor was applied.

\begin{itemize}
\item \emph{ECAPA-TDNN}: The ECAPA-TDNN speaker encoder was trained on the VoxCeleb1 \cite{nagrani2017voxceleb} and VoxCeleb2 \cite{chung2018voxceleb2} datasets, utilizing the open-source recipe ASV-Subtools\footnote{https://github.com/Snowdar/asv-subtools}.

\item \emph{FACodec}: The timbre encoder in the open-source FACodec \cite{facodec} model\footnote{https://github.com/lifeiteng/naturalspeech3\_facodec} was used. It was trained on a 60K-hour Libri-light dataset \cite{kahn2020libri}.

\end{itemize}

The Diff-Net module comprised two fully-connected layers. The output size of the first layer was 128. The model was trained for 10 epochs, with a batch size of 16. The learning rate for extracting embeddings using the ECAPA-TDNN pre-trained model was 0.00005, whereas the learning rate for extracting embeddings with the FACodec pre-trained model was 0.000025.

% During the training, all attributes listed in Table \ref{table1} were used. Additionally, all speech samples were downsampled to 16 kHz. For each annotation group, 20 different speech samples were selected for each speaker. Two types of embeddings were extracted from the pre-trained ECAPA-TDNN and FACodec models, with the latter's embeddings specifically derived from the Timbre Extractor module. To ensure proper convergence of both models, different learning rates were applied during training. The model was trained for 10 epochs, with each epoch consisting of several batches, each having a batch size of 16.

\subsection{Experimental Results}

 In the verification evaluation, given an ordered utterance pair $\langle{\mathcal O}_{\rm A},{\mathcal O}_{\rm B}\rangle$ and a designated descriptor v, the output value of the node corresponding v was used as the confidence score for EER computation. In the recognition evaluation, a threshold of 0.5 was employed, with values greater than 0.5 decided as true and values less than 0.5 decided as false. The models were evaluated in the three scenarios as described in Section \ref{sec:eval scenarios}, including unseen, seen-speaker, and seen-speaker-pair. The results obtained on the ECAPA-TDNN and FACodec speaker encoders in the three scenarios can be found in Tables \ref{tab: result of the unseen set}, \ref{tab: result of the seen set}, and \ref{tab: result of the seen speaker pair set}, respectively.

\paragraph{ECAPA-TDNN:}
%unseen speaker test male分析
The results on the unseen test set indicated that when the ECAPA-TDNN model was used as a pre-trained speaker encoder for the vTAD task, its accuracy was generally lower for male speakers, with an average accuracy of only 73.41\% and an average EER as high as 26.11\%. 
%unseen speaker test female分析
However, for female speakers, the model demonstrated high accuracy and low EER in distinguishing the ``Coarse'' and ``Slim'' attributes, with accuracy of 91.90\% and 91.50\%, and EERs of 9.26\% and 7.73\%, respectively. In contrast, the performance in distinguishing the ``Bright'' and ``Thin'' attributes was much poorer, with accuracy of only 50.36\% and 46.88\%, and EERs of 49.34\% and 53.10\%, respectively. 
% 其他两个测试集分析
Further comparisons on different test sets showed that the model achieved average accuracies exceeding 90\% and average EERs below 10\% on both the seen-speaker test set and the seen-speaker-pair set. These results were significantly better than those on the unseen test set. 
This observation highlighted the limitations of the ECAPA-TDNN-based approach in terms of generalization capability.

\paragraph{FACodec:}
%对比ECAPA
When using FACodec as the pretrained speaker encoder to build the model, it outperformed the ECAPA-TDNN-based model in most of the vTAD tasks. The average accuracy on the unseen test set was 91.79\% for male speakers and 89.74\% for female speakers, while the average EER was 8.41\% for males and 10.21\% for females.
%另外两个数据集
Additionally, the model achieved average accuracies of over 90\% on both the seen-speaker test set and the seen-speaker-pair set, with average EERs below 10\%, which are slightly higher than the results on the unseen test set.
These findings strongly demonstrated that the FACodec method offered a performance advantage and exhibited good generalization ability, showing strong effectiveness and reliability in the vTAD task.

% 结论
\section{Conclusions}

This paper introduced a novel task, voice timbre attribute detection, which involves comparing the intensity between two speech utterances in a specific timbre descriptor. It is aimed at explaining the timbre of voice, thereby enhancing the understanding of timbre. In this study, a framework was proposed, built upon the speaker embeddings extracted from a pair of speech utterances. The investigation was conducted on the VCTK-RVA dataset. The experiments examined two speaker encoders for speaker embedding extraction: one derived from the ECAPA-TDNN model and the other from the timbre extractor of the FACodec model. The performance metrics obtained from both verification and recognition evaluations indicate that the intensity differences between two utterances are detectable in the timbre descriptors. Between the ECAPA-TDNN and FACodec speaker encoders, the former demonstrated superior performance in the seen scenario, where the testing speakers were included in the training set. The latter exhibited better capability in the unseen scenario, where the testing speakers were not part of the training, indicating enhanced generalization capability.

\appendix
\clearpage

\section{Tables}

\begin{table}[ht]
	\centering
	\caption{The statistics of the male and female speakers in the training set. The number of ordered speaker pairs (\emph{\#Pairs}) and the number of speakers (\emph{\#Speakers}) are presented for each descriptor (\emph{Descr.}).}
    {
	\begin{tabular}{lcc lcc}
		\toprule[1pt]
        \multicolumn{3}{c}{\textbf{Male}} & \multicolumn{3}{c}{\textbf{Female}}\\
		\textbf{Descr.} & \textbf{\#Pairs} & \textbf{\#Speakers} & \textbf{Descr.} & \textbf{\#Pair} & \textbf{\#Speakers} \\ 
		\cmidrule(lr){1-3} \cmidrule(lr){4-6}
		Bright(\begin{CJK*}{UTF8}{gbsn}明亮\end{CJK*}) & 182 & 29 & Bright(\begin{CJK*}{UTF8}{gbsn}明亮\end{CJK*}) & 428 & 49 \\
		Thin(\begin{CJK*}{UTF8}{gbsn}单薄\end{CJK*}) & 82 & 29 & Thin(\begin{CJK*}{UTF8}{gbsn}单薄\end{CJK*}) & 351 & 49 \\
		Low(\begin{CJK*}{UTF8}{gbsn}低沉\end{CJK*}) & 70 & 26 & Low(\begin{CJK*}{UTF8}{gbsn}低沉\end{CJK*}) & 191 & 48 \\
		Magnetic(\begin{CJK*}{UTF8}{gbsn}磁性\end{CJK*}) & 60 & 29 & Magnetic(\begin{CJK*}{UTF8}{gbsn}磁性\end{CJK*}) & 44 & 38 \\
		Coarse(\begin{CJK*}{UTF8}{gbsn}粗\end{CJK*}) & 64 & 27 & Coarse(\begin{CJK*}{UTF8}{gbsn}粗\end{CJK*}) & 382 & 49 \\
		Slim(\begin{CJK*}{UTF8}{gbsn}细\end{CJK*}) & 56 & 27 & Slim(\begin{CJK*}{UTF8}{gbsn}细\end{CJK*}) & 373 & 49 \\
		Muddy(\begin{CJK*}{UTF8}{gbsn}浑浊\end{CJK*}) & 54 & 27 & Muddy(\begin{CJK*}{UTF8}{gbsn}浑浊\end{CJK*}) & 106 & 44 \\
		Muffled(\begin{CJK*}{UTF8}{gbsn}沉闷\end{CJK*}) & 53 & 25 & Muffled(\begin{CJK*}{UTF8}{gbsn}沉闷\end{CJK*}) & 7 & 14 \\
		Pure(\begin{CJK*}{UTF8}{gbsn}干净\end{CJK*}) & 46 & 23 & Pure(\begin{CJK*}{UTF8}{gbsn}干净\end{CJK*}) & 196 & 47 \\
		Soft(\begin{CJK*}{UTF8}{gbsn}柔和\end{CJK*}) & 36 & 24 & Soft(\begin{CJK*}{UTF8}{gbsn}柔和\end{CJK*}) & 7 & 14 \\
		Flat(\begin{CJK*}{UTF8}{gbsn}平淡\end{CJK*}) & 30 & 23 & Flat(\begin{CJK*}{UTF8}{gbsn}平淡\end{CJK*}) & 59 & 36 \\
		Hoarse(\begin{CJK*}{UTF8}{gbsn}沙哑\end{CJK*}) & 26 & 25 & Hoarse(\begin{CJK*}{UTF8}{gbsn}沙哑\end{CJK*}) & 126 & 49 \\
		Rich(\begin{CJK*}{UTF8}{gbsn}厚实\end{CJK*}) & 24 & 22 & Rich(\begin{CJK*}{UTF8}{gbsn}厚实\end{CJK*}) & 159 & 47 \\
		Shriveled(\begin{CJK*}{UTF8}{gbsn}干瘪\end{CJK*}) & 23 & 21 & Shriveled(\begin{CJK*}{UTF8}{gbsn}干瘪\end{CJK*}) & 19 & 22 \\
		Round(\begin{CJK*}{UTF8}{gbsn}圆润\end{CJK*}) & 14 & 14 & Round(\begin{CJK*}{UTF8}{gbsn}圆润\end{CJK*}) & 35 & 31 \\
		Transparent(\begin{CJK*}{UTF8}{gbsn}通透\end{CJK*}) & 10 & 15 & Transparent(\begin{CJK*}{UTF8}{gbsn}通透\end{CJK*}) & 2 & 4 \\
		Husky(\begin{CJK*}{UTF8}{gbsn}干哑\end{CJK*}) & 10 & 15 & Shrill(\begin{CJK*}{UTF8}{gbsn}尖锐\end{CJK*}) & 69 & 45 \\
		\bottomrule[1pt]
	\end{tabular}
    }
	\label{tab: training spkr stats}
\end{table}

%table3  the unseen speaker test set
\begin{table}[ht]
	\centering
		\caption{The statistics of the male and female speakers in the unseen test set. The number of ordered speaker pairs (\emph{\#Pairs}) and the number of speakers (\emph{\#Speakers}) are presented for each descriptor (\emph{Descr.}).}
{
	\begin{tabular}{lcc lcc}
		\toprule[1pt]
        \multicolumn{3}{c}{\textbf{Male}} & \multicolumn{3}{c}{\textbf{Female}}\\
		\textbf{Descr.} & \textbf{\#Pairs} & \textbf{\#Speakers} & \textbf{Descr.} & \textbf{\#Pairs} & \textbf{\#Speakers} \\ 
		\cmidrule(lr){1-3} \cmidrule(lr){4-6}
		Bright(\begin{CJK*}{UTF8}{gbsn}明亮\end{CJK*}) & 34 & 20     & Bright(\begin{CJK*}{UTF8}{gbsn}明亮\end{CJK*})  & 35 & 40 \\
        Thin(\begin{CJK*}{UTF8}{gbsn}单薄\end{CJK*}) & 29 & 10 
        & Thin(\begin{CJK*}{UTF8}{gbsn}单薄\end{CJK*}) & 28 & 35  \\
        Low(\begin{CJK*}{UTF8}{gbsn}低沉\end{CJK*}) & 13 & 10 
        & Low(\begin{CJK*}{UTF8}{gbsn}低沉\end{CJK*}) & 15 & 10  \\
        Magnetic(\begin{CJK*}{UTF8}{gbsn}磁性\end{CJK*}) & 17 & 10 
        & Coarse(\begin{CJK*}{UTF8}{gbsn}粗\end{CJK*}) & 26 & 40  \\
        Pure(\begin{CJK*}{UTF8}{gbsn}干净\end{CJK*}) & 6 & 5 
        & Slim(\begin{CJK*}{UTF8}{gbsn}细\end{CJK*}) & 26 & 40  \\
		\bottomrule[1pt]
	\end{tabular}
        }
	\label{tab: unseen-spkr eval stats}
\end{table}

%table4  the seen speaker test set
\begin{table}[ht]
	\centering
		\caption{The statistics of the male and female speakers in the seen test set. The number of ordered speaker pairs (\emph{\#Pairs}) and the number of speakers (\emph{\#Speakers}) are presented for each descriptor (\emph{Descr.}).}
{
	\begin{tabular}{lcc lcc}
		\toprule[1pt]
        \multicolumn{3}{c}{\textbf{Male}} & \multicolumn{3}{c}{\textbf{Female}}\\
		\textbf{Descr.} & \textbf{\#Pairs} & \textbf{\#Speakers} & \textbf{Descr.} & \textbf{\#Pair} & \textbf{\#Speakers} \\ 
		\cmidrule(lr){1-3} \cmidrule(lr){4-6}
		Bright(\begin{CJK*}{UTF8}{gbsn}明亮\end{CJK*}) & 20 & 21     & Bright(\begin{CJK*}{UTF8}{gbsn}明亮\end{CJK*})  & 40 & 39 \\
        Thin(\begin{CJK*}{UTF8}{gbsn}单薄\end{CJK*}) & 10 & 12 
        & Thin(\begin{CJK*}{UTF8}{gbsn}单薄\end{CJK*}) & 35 & 33  \\
        Low(\begin{CJK*}{UTF8}{gbsn}低沉\end{CJK*}) & 10 & 13 
        & Low(\begin{CJK*}{UTF8}{gbsn}低沉\end{CJK*}) & 20 & 26  \\
        Magnetic(\begin{CJK*}{UTF8}{gbsn}磁性\end{CJK*}) & 10 & 13 
        & Coarse(\begin{CJK*}{UTF8}{gbsn}粗\end{CJK*}) & 20 & 26  \\
        Pure(\begin{CJK*}{UTF8}{gbsn}干净\end{CJK*}) & 10 & 13 
        & Slim(\begin{CJK*}{UTF8}{gbsn}细\end{CJK*}) & 20 & 26  \\
		\bottomrule[1pt]
	\end{tabular}
        }
	\label{tab: seen-spkr eval stats}
\end{table}

\begin{table}[ht]
	\centering
	\caption{Evaluation results of the vTAD model on the unseen test set. The row \emph{Avg} is obtained by averaging the results across all the descriptors for each metric.}
	\begin{tabular}{c|ccc|ccc}

		\toprule[1pt]
		 & \multicolumn{3}{c|}{\textbf{Male}} & \multicolumn{3}{c}{\textbf{Female}} \\ 
	  \cmidrule(lr){2-4} \cmidrule(lr){5-7}
	\textbf{Model} &Attr. & ACC (\%) & EER (\%) & Attr. & ACC (\%) & EER (\%) \\ 
		\midrule
        
		& Bright(\begin{CJK*}{UTF8}{gbsn}明亮\end{CJK*})    & 66.55 & 34.81 & Bright(\begin{CJK*}{UTF8}{gbsn}明亮\end{CJK*})  & 50.36 & 49.34 \\
		& Thin(\begin{CJK*}{UTF8}{gbsn}单薄\end{CJK*})      & 72.19 & 27.40 & Thin(\begin{CJK*}{UTF8}{gbsn}单薄\end{CJK*})    & 46.88 & 53.10 \\
	  & Low(\begin{CJK*}{UTF8}{gbsn}低沉\end{CJK*})       & 80.90 & 18.77 & Low(\begin{CJK*}{UTF8}{gbsn}低沉\end{CJK*})     & 67.52 & 33.42 \\
		\textbf{ ECAPA-TDNN}  & Magnetic(\begin{CJK*}{UTF8}{gbsn}磁性\end{CJK*})   & 79.04 & 17.59 & Coarse(\begin{CJK*}{UTF8}{gbsn}粗\end{CJK*})  & 91.90 & 9.26  \\
		& Pure(\begin{CJK*}{UTF8}{gbsn}干净\end{CJK*})      & 68.38 & 32.00 & Slim(\begin{CJK*}{UTF8}{gbsn}细\end{CJK*})    & 91.50 & 7.73  \\
		\cmidrule(lr){2-7}
        & Avg    &73.41  & 26.11 & Avg   & 69.63 & 30.57 \\

		\midrule
        
		& Bright(\begin{CJK*}{UTF8}{gbsn}明亮\end{CJK*})      & 93.60 & 6.08  & Bright(\begin{CJK*}{UTF8}{gbsn}明亮\end{CJK*})  & 88.41 & 11.42 \\
		& Thin(\begin{CJK*}{UTF8}{gbsn}单薄\end{CJK*})        & 94.84 & 4.67  & Thin(\begin{CJK*}{UTF8}{gbsn}单薄\end{CJK*})    & 89.61 & 10.32 \\
		 & Low(\begin{CJK*}{UTF8}{gbsn}低沉\end{CJK*})         & 91.79 & 10.15 & Low(\begin{CJK*}{UTF8}{gbsn}低沉\end{CJK*})     & 86.23 & 13.56 \\
		\textbf{ FACodec} & Magnetic(\begin{CJK*}{UTF8}{gbsn}磁性\end{CJK*})    & 98.31 & 1.96  & Coarse(\begin{CJK*}{UTF8}{gbsn}粗\end{CJK*})  & 90.92 & 9.05  \\
		& Pure(\begin{CJK*}{UTF8}{gbsn}干净\end{CJK*})        & 80.42 & 19.17 & Slim(\begin{CJK*}{UTF8}{gbsn}细\end{CJK*})    & 93.51 & 6.68  \\
		\cmidrule(lr){2-7}
        & Avg    &91.79  & 8.41  & Avg   & 89.74 & 10.21 \\

		\bottomrule[1pt]
	\end{tabular}
	\label{tab: result of the unseen set}
\end{table}

\begin{table}[ht]
	\centering
	\caption{Evaluation results of the vTAD model on the seen-speaker test set. The row \emph{Avg} is obtained by averaging the results across all the descriptors for each metric.}
	\begin{tabular}{c|ccc|ccc}

		\toprule[1pt]
		 & \multicolumn{3}{c|}{\textbf{Male}} & \multicolumn{3}{c}{\textbf{Female}} \\ 
	  \cmidrule(lr){2-4} \cmidrule(lr){5-7}
		 \textbf{Model} &Attr. & ACC (\%) & EER (\%) & Attr. & ACC (\%) & EER (\%) \\ 
		\midrule
		& Bright(\begin{CJK*}{UTF8}{gbsn}明亮\end{CJK*})      & 95.99 & 4.45  & Bright(\begin{CJK*}{UTF8}{gbsn}明亮\end{CJK*})  & 90.44 & 9.66 \\
		& Thin(\begin{CJK*}{UTF8}{gbsn}单薄\end{CJK*})        & 96.40 & 3.50  & Thin(\begin{CJK*}{UTF8}{gbsn}单薄\end{CJK*})    & 91.03 & 8.55 \\
		 & Low(\begin{CJK*}{UTF8}{gbsn}低沉\end{CJK*})         & 99.20 & 0.87 & Low(\begin{CJK*}{UTF8}{gbsn}低沉\end{CJK*})     & 97.86 & 2.30 \\
		\textbf{ ECAPA-TDNN} & Magnetic(\begin{CJK*}{UTF8}{gbsn}磁性\end{CJK*})    & 97.50 & 2.60  & Coarse(\begin{CJK*}{UTF8}{gbsn}粗\end{CJK*})  & 93.97 & 6.58  \\
		& Pure(\begin{CJK*}{UTF8}{gbsn}干净\end{CJK*})        & 84.97 & 15.07 & Slim(\begin{CJK*}{UTF8}{gbsn}细\end{CJK*})    & 97.76 & 1.57  \\
		\cmidrule(lr){2-7}
        & Avg    &94.81  & 5.30  & Avg   & 94.21 & 5.73 \\

		\midrule
        
		& Bright(\begin{CJK*}{UTF8}{gbsn}明亮\end{CJK*})      & 96.78 & 3.00  & Bright(\begin{CJK*}{UTF8}{gbsn}明亮\end{CJK*})  & 90.71 & 9.29 \\
		& Thin(\begin{CJK*}{UTF8}{gbsn}单薄\end{CJK*})        & 90.97 & 8.37  & Thin(\begin{CJK*}{UTF8}{gbsn}单薄\end{CJK*})    & 93.40 & 6.64 \\
		 & Low(\begin{CJK*}{UTF8}{gbsn}低沉\end{CJK*})         & 97.02 & 3.17 & Low(\begin{CJK*}{UTF8}{gbsn}低沉\end{CJK*})     & 98.70 & 1.42 \\
		\textbf{ FACodec} & Magnetic(\begin{CJK*}{UTF8}{gbsn}磁性\end{CJK*})    & 97.97 & 2.90  & Coarse(\begin{CJK*}{UTF8}{gbsn}粗\end{CJK*})  & 88.29 & 11.25 \\
		& Pure(\begin{CJK*}{UTF8}{gbsn}干净\end{CJK*})        & 81.08 & 16.10 & Slim(\begin{CJK*}{UTF8}{gbsn}细\end{CJK*})    & 97.65 & 2.38 \\
		\cmidrule(lr){2-7}
        & Avg    &92.77  & 6.71  & Avg   & 93.75 & 6.20 \\  

		\bottomrule[1pt]
	\end{tabular}
	\label{tab: result of the seen set}
\end{table}

\begin{table}[ht]
	\centering
	\caption{Evaluation results of the vTAD model on the seen-speaker-pair test set. The row \emph{Avg} is obtained by averaging the results across all the descriptors for each metric.}
    \resizebox{0.97\textwidth}{!}{
	\begin{tabular}{c|ccc|ccc}
		\toprule[1pt]
		& \multicolumn{3}{c|}{\textbf{Male}} & \multicolumn{3}{c}{\textbf{Female}} \\ 
		\cmidrule(lr){2-4} \cmidrule(lr){5-7}
		\textbf{Model} &Attr. & ACC (\%) & EER (\%) & Attr. & ACC (\%) & EER (\%) \\ 
		\midrule
        
& Bright(\begin{CJK*}{UTF8}{gbsn}明亮\end{CJK*})      & 95.90  & 4.47  & Bright(\begin{CJK*}{UTF8}{gbsn}明亮\end{CJK*})      & 96.89  & 3.18 \\
& Thin(\begin{CJK*}{UTF8}{gbsn}单薄\end{CJK*})        & 97.18  & 2.76  & Thin(\begin{CJK*}{UTF8}{gbsn}单薄\end{CJK*})        & 98.09  & 1.82 \\
& Low(\begin{CJK*}{UTF8}{gbsn}低沉\end{CJK*})         & 99.41  & 0.67  & Coarse(\begin{CJK*}{UTF8}{gbsn}粗\end{CJK*})        & 97.82  & 2.45 \\
& Magnetic(\begin{CJK*}{UTF8}{gbsn}磁性\end{CJK*})    & 99.47  & 0.47  & Slim(\begin{CJK*}{UTF8}{gbsn}细\end{CJK*})          & 96.84  & 2.69 \\
& Coarse(\begin{CJK*}{UTF8}{gbsn}粗\end{CJK*})        & 99.92  & 0.08  & Pure(\begin{CJK*}{UTF8}{gbsn}干净\end{CJK*})        & 95.61  & 4.57 \\
& Slim(\begin{CJK*}{UTF8}{gbsn}细\end{CJK*})          & 99.46  & 0.52  & Low(\begin{CJK*}{UTF8}{gbsn}低沉\end{CJK*})         & 97.37  & 2.72 \\
& Muddy(\begin{CJK*}{UTF8}{gbsn}浑浊\end{CJK*})       & 98.67  & 1.16  & Hoarse(\begin{CJK*}{UTF8}{gbsn}沙哑\end{CJK*})      & 98.45  & 1.53 \\
& Muffled(\begin{CJK*}{UTF8}{gbsn}沉闷\end{CJK*})     & 98.92  & 1.06  & Rich(\begin{CJK*}{UTF8}{gbsn}厚实\end{CJK*})        & 97.61  & 2.76 \\
\textbf{ ECAPA-TDNN}& Pure(\begin{CJK*}{UTF8}{gbsn}干净\end{CJK*})        & 99.33  & 0.87  & Muddy(\begin{CJK*}{UTF8}{gbsn}浑浊\end{CJK*})       & 93.64  & 6.73 \\
& Soft(\begin{CJK*}{UTF8}{gbsn}柔和\end{CJK*})        & 98.53  & 1.44  & Flat(\begin{CJK*}{UTF8}{gbsn}平淡\end{CJK*})        & 94.58  & 6.42 \\
& Flat(\begin{CJK*}{UTF8}{gbsn}平淡\end{CJK*})        & 94.20  & 5.73  & Shrill(\begin{CJK*}{UTF8}{gbsn}尖锐\end{CJK*})      & 99.86  & 0.12 \\
& Hoarse(\begin{CJK*}{UTF8}{gbsn}沙哑\end{CJK*})      & 99.85  & 0.21  & Magnetic(\begin{CJK*}{UTF8}{gbsn}磁性\end{CJK*})    & 99.20  & 1.12 \\
& Rich(\begin{CJK*}{UTF8}{gbsn}厚实\end{CJK*})        & 99.62  & 0.61  & Round(\begin{CJK*}{UTF8}{gbsn}圆润\end{CJK*})       & 98.97  & 1.03 \\
& Shriveled(\begin{CJK*}{UTF8}{gbsn}干瘪\end{CJK*})   & 97.96  & 1.04  & Shriveled(\begin{CJK*}{UTF8}{gbsn}干瘪\end{CJK*})   & 98.05  & 0.70 \\
& Round(\begin{CJK*}{UTF8}{gbsn}圆润\end{CJK*})       & 99.79  & 0.19  & Soft(\begin{CJK*}{UTF8}{gbsn}柔和\end{CJK*})        & 98.96  & 0.00 \\
& Transparent(\begin{CJK*}{UTF8}{gbsn}通透\end{CJK*}) & 99.50  & 1.07  & Muffled(\begin{CJK*}{UTF8}{gbsn}沉闷\end{CJK*})     & 100.00 & 0.00 \\
& Husky(\begin{CJK*}{UTF8}{gbsn}干哑\end{CJK*})       & 100.00  & 0.00  & Transparent(\begin{CJK*}{UTF8}{gbsn}通透\end{CJK*}) & 100.00 & 0.00 \\
\cmidrule(lr){2-7}
& Avg    & 98.69 & 1.31 & Avg   & 97.81 & 2.23 \\

		\midrule
& Bright(\begin{CJK*}{UTF8}{gbsn}明亮\end{CJK*})      & 94.71  & 5.28  & Bright(\begin{CJK*}{UTF8}{gbsn}明亮\end{CJK*})      & 96.36  & 3.67 \\
& Thin(\begin{CJK*}{UTF8}{gbsn}单薄\end{CJK*})        & 97.91  & 1.89  & Thin(\begin{CJK*}{UTF8}{gbsn}单薄\end{CJK*})        & 98.73  & 1.26 \\
& Low(\begin{CJK*}{UTF8}{gbsn}低沉\end{CJK*})         & 99.10  & 1.01  & Coarse(\begin{CJK*}{UTF8}{gbsn}粗\end{CJK*})        & 94.89  & 4.77 \\
& Magnetic(\begin{CJK*}{UTF8}{gbsn}磁性\end{CJK*})    & 99.20  & 0.83  & Slim(\begin{CJK*}{UTF8}{gbsn}细\end{CJK*})          & 95.54  & 4.83 \\
& Coarse(\begin{CJK*}{UTF8}{gbsn}粗\end{CJK*})        & 99.61  & 0.46  & Pure(\begin{CJK*}{UTF8}{gbsn}干净\end{CJK*})        & 90.82  & 8.11 \\
& Slim(\begin{CJK*}{UTF8}{gbsn}细\end{CJK*})          & 98.77  & 0.83  & Low(\begin{CJK*}{UTF8}{gbsn}低沉\end{CJK*})         & 95.34  & 4.52 \\
& Muddy(\begin{CJK*}{UTF8}{gbsn}浑浊\end{CJK*})       & 97.24  & 3.53  & Hoarse(\begin{CJK*}{UTF8}{gbsn}沙哑\end{CJK*})      & 96.88  & 3.54 \\
& Muffled(\begin{CJK*}{UTF8}{gbsn}沉闷\end{CJK*})     & 97.28  & 2.69  & Rich(\begin{CJK*}{UTF8}{gbsn}厚实\end{CJK*})        & 95.82  & 3.87 \\
\textbf{ FACodec} & Pure(\begin{CJK*}{UTF8}{gbsn}干净\end{CJK*})        & 95.20  & 3.33  & Muddy(\begin{CJK*}{UTF8}{gbsn}浑浊\end{CJK*})       & 87.66  & 13.76 \\
& Soft(\begin{CJK*}{UTF8}{gbsn}柔和\end{CJK*})        & 95.22  & 5.15  & Flat(\begin{CJK*}{UTF8}{gbsn}平淡\end{CJK*})        & 90.92  & 8.95 \\
& Flat(\begin{CJK*}{UTF8}{gbsn}平淡\end{CJK*})        & 90.77  & 7.78  & Shrill(\begin{CJK*}{UTF8}{gbsn}尖锐\end{CJK*})      & 99.62  & 0.39 \\
& Hoarse(\begin{CJK*}{UTF8}{gbsn}沙哑\end{CJK*})      & 99.62  & 0.56  & Magnetic(\begin{CJK*}{UTF8}{gbsn}磁性\end{CJK*})    & 97.39  & 2.67 \\
& Rich(\begin{CJK*}{UTF8}{gbsn}厚实\end{CJK*})        & 96.96  & 3.50  & Round(\begin{CJK*}{UTF8}{gbsn}圆润\end{CJK*})       & 94.71  & 5.64 \\
& Shriveled(\begin{CJK*}{UTF8}{gbsn}干瘪\end{CJK*})   & 94.70  & 5.16  & Shriveled(\begin{CJK*}{UTF8}{gbsn}干瘪\end{CJK*})   & 95.58  & 4.77 \\
& Round(\begin{CJK*}{UTF8}{gbsn}圆润\end{CJK*})       & 92.93  & 5.43  & Soft(\begin{CJK*}{UTF8}{gbsn}柔和\end{CJK*})        & 96.43  & 3.24 \\
& Transparent(\begin{CJK*}{UTF8}{gbsn}通透\end{CJK*}) & 97.00  & 3.07  & Muffled(\begin{CJK*}{UTF8}{gbsn}沉闷\end{CJK*})     & 96.75  & 5.00 \\
& Husky(\begin{CJK*}{UTF8}{gbsn}干哑\end{CJK*})       & 100.00  & 0.00  & Transparent(\begin{CJK*}{UTF8}{gbsn}通透\end{CJK*}) & 83.00 & 11.33 \\
\cmidrule(lr){2-7}
& Avg    & 96.84 & 2.97 & Avg   & 94.50 & 5.31 \\

		\bottomrule[1pt]
	\end{tabular}
    }
	\label{tab: result of the seen speaker pair set}
\end{table}

\clearpage
\bibliographystyle{IEEEbib}
\bibliography{sample}

\begin{thebibliography}{10}

\bibitem{x-vector}
David Snyder, Daniel Garcia-Romero, Daniel Povey, and Sanjeev Khudanpur,
\newblock ``Deep neural network embeddings for text-independent speaker verification.,''
\newblock in {\em Proc. Interspeech}, 2017, pp. 999--1003.

\bibitem{desplanques2020ecapa}
Brecht Desplanques, Jenthe Thienpondt, and Kris Demuynck,
\newblock ``{ECAPA-TDNN}: Emphasized channel attention, propagation and aggregation in {TDNN} based speaker verification,''
\newblock in {\em Proc. Interspeech}, 2020, pp. 3830--3834.

\bibitem{ExplainableASV}
Xiaoliang Wu, Chau Luu, Peter Bell, and Ajitha Rajan,
\newblock ``Explainable attribute-based speaker verification,''
\newblock {\em CoRR}, vol. abs/2405.19796, 2024.

\bibitem{li2022recent}
Jinyu Li et~al.,
\newblock ``Recent advances in end-to-end automatic speech recognition,''
\newblock {\em APSIPA Transactions on Signal and Information Processing}, vol. 11, no. 1, 2022.

\bibitem{yourtts}
Edresson Casanova, Julian Weber, Christopher~D Shulby, et~al.,
\newblock ``{Y}our{TTS}: Towards zero-shot multi-speaker {TTS} and zero-shot voice conversion for everyone,''
\newblock in {\em Proc. International Conference on Machine Learning}, 2022, vol. 162, pp. 2709--2720.

\bibitem{voice_conversion_overview}
B.~Sisman, J.~Yamagishi, S.~King, and H.~Li,
\newblock ``An overview of voice conversion and its challenges: from statistical modeling to deep learning,''
\newblock {\em IEEE/ACM Transactions on Audio, Speech, and Language Processing}, vol. 29, pp. 132--157, 2021.

\bibitem{wang2024async}
Rui Wang, Liping Chen, Kong~Aik Lee, and Zhen-Hua Ling,
\newblock ``Asynchronous voice anonymization using adversarial perturbation on speaker embedding,''
\newblock in {\em Proc. Interspeech}, 2024, pp. 4443--4447.

\bibitem{DBLP:journals/corr/abs-2404-08857}
Zhengyan Sheng, Yang Ai, Li{-}Juan Liu, Jia Pan, and Zhen{-}Hua Ling,
\newblock ``Voice attribute editing with text prompt,''
\newblock {\em CoRR}, vol. abs/2404.08857, 2024.

\bibitem{facodec}
Zeqian Ju, Yuancheng Wang, Kai Shen, et~al.,
\newblock ``Naturalspeech 3: Zero-shot speech synthesis with factorized codec and diffusion models,''
\newblock in {\em Proc. {ICML}}, 2024.

\bibitem{vctk-rva}
Zheng-Yan Sheng, Li-Juan Liu, Yang Ai, Jia Pan, and Zhen-Hua Ling,
\newblock ``Voice attribute editing with text prompt,''
\newblock {\em IEEE Transactions on Audio, Speech and Language Processing}, vol. 33, pp. 1641--1652, 2025.

\bibitem{resnet}
Hossein Zeinali, Shuai Wang, Anna Silnova, et~al.,
\newblock ``{BUT} system description to {VoxCeleb} speaker recognition challenge 2019,''
\newblock {\em arXiv preprint arXiv:1910.12592}, 2019.

\bibitem{nagrani2017voxceleb}
Arsha Nagrani, Joon~Son Chung, and Andrew Zisserman,
\newblock ``Voxceleb: A large-scale speaker identification dataset,''
\newblock {\em arXiv preprint arXiv:1706.08612}, 2017.

\bibitem{chung2018voxceleb2}
Joon~Son Chung, Arsha Nagrani, and Andrew Zisserman,
\newblock ``Voxceleb2: Deep speaker recognition,''
\newblock {\em arXiv preprint arXiv:1806.05622}, 2018.

\bibitem{kahn2020libri}
Jacob Kahn, Morgane Riviere, Weiyi Zheng, et~al.,
\newblock ``Libri-light: A benchmark for {ASR} with limited or no supervision,''
\newblock in {\em {Proc. ICASSP}}. IEEE, 2020, pp. 7669--7673.

\end{thebibliography}

\end{document}